\documentclass[10pt,letterpaper]{article}

\usepackage[T1]{fontenc}
\usepackage[utf8]{inputenc}
\usepackage{newtxtext,newtxmath}
\usepackage{microtype}
\usepackage[letterpaper,top=0.68in,bottom=0.72in,left=0.74in,right=0.74in,
            headheight=15pt,headsep=16pt,footskip=26pt]{geometry}
\usepackage{graphicx}
\usepackage{xcolor}
\usepackage{tikz}
\usetikzlibrary{arrows.meta,positioning}
\usepackage{booktabs}
\usepackage{float}
\usepackage{array}
\usepackage{enumitem}
\usepackage{amsmath}
\usepackage{titlesec}
\usepackage{fancyhdr}
\usepackage[font=small,labelfont={bf,rm},labelsep=period]{caption}
\usepackage[numbers,square,sort&compress]{natbib}
\usepackage[most]{tcolorbox}
\usepackage{hyperref}

\definecolor{Cobalt}{HTML}{063166}
\definecolor{Teal}{HTML}{5384B2}
\definecolor{Ink}{HTML}{172539}
\definecolor{Subtle}{HTML}{58697C}
\definecolor{SoftRule}{HTML}{D9E3ED}
\definecolor{Pale}{HTML}{F1F6FB}
\definecolor{PaleBlue}{HTML}{E7F0F8}
\hypersetup{colorlinks=true,linkcolor=Cobalt,citecolor=Cobalt,urlcolor=Cobalt,
  pdftitle={APCL Research White Paper Template},
  pdfauthor={Advanced Propulsion Concepts Lab (APCL)}}

\newcommand{\labname}{Advanced Propulsion Concepts Lab}
\newcommand{\papertitle}{Flow Structure in Idealized Rotating Detonation \\Combustor with Supersonic Inflow}
\newcommand{\paperauthors}{%
  Ruben A. Fernandez\textsuperscript{1,*},\quad
  Venkat Raman\textsuperscript{1}}
\newcommand{\paperdate}{November 2026}
\newcommand{\paperaffiliations}{%
  \textsuperscript{1}\labname{} (APCL), Department of Aerospace Engineering, University of Michigan, Ann Arbor, 48109, USA\\[2pt]}

\newcommand{\papercorrespondence}{Ruben A. Fernandez (\href{mailto:author@example.org}{rubenf@umich.edu})}

\definecolor{mywhite}{HTML}{FFFFFF}
\newcommand{\whitecircle}{%
  \tikz[baseline=-0.5ex]\node[circle,fill=mywhite,draw=black,inner sep=3pt]{};%
}

\definecolor{myblack}{HTML}{000000}
\newcommand{\blackcircle}{%
  \tikz[baseline=-0.5ex]\node[circle,fill=myblack,draw=black,inner sep=3pt]{};%
}

\definecolor{mybrightblue}{HTML}{0000FF}
\newcommand{\brightbluecircle}{%
  \tikz[baseline=-0.5ex]\node[circle,fill=mybrightblue,draw=black,inner sep=3pt]{};%
}

\definecolor{myblue}{HTML}{1f77b4}
\newcommand{\bluecircle}{%
  \tikz[baseline=-0.5ex]\node[circle,fill=myblue,draw=black,inner sep=3pt]{};%
}

\definecolor{myorange}{HTML}{ff7f0e}
\newcommand{\orangecircle}{%
  \tikz[baseline=-0.5ex]\node[circle,fill=myorange,draw=black,inner sep=3pt]{};%
}

\definecolor{mygreen}{HTML}{2ca02c}
\newcommand{\greencircle}{%
  \tikz[baseline=-0.5ex]\node[circle,fill=mygreen,draw=black,inner sep=3pt]{};%
}

\definecolor{myred}{HTML}{d62728}
\newcommand{\redcircle}{%
  \tikz[baseline=-0.5ex]\node[circle,fill=myred,draw=black,inner sep=3pt]{};%
}

\definecolor{mypurple}{HTML}{9467bd}
\newcommand{\purplecircle}{%
  \tikz[baseline=-0.5ex]\node[circle,fill=mypurple,draw=black,inner sep=3pt]{};%
}

\color{Ink}
\setlist[itemize]{topsep=3pt,itemsep=3pt,parsep=0pt,leftmargin=1.35em,
  label={\textcolor{Cobalt}{\large\textbullet}}}
\titleformat{\section}{\rmfamily\bfseries\large\color{Ink}}
  {\textcolor{Cobalt}{\thesection}}{0.70em}{}
\titleformat{\subsection}{\rmfamily\bfseries\normalsize\color{Ink}}
  {\textcolor{Cobalt}{\thesubsection}}{0.70em}{}
\titlespacing*{\section}{0pt}{17pt plus 2pt minus 1pt}{6pt}
\titlespacing*{\subsection}{0pt}{11pt plus 1pt minus 1pt}{4pt}
\fancypagestyle{apcl}{%
  \fancyhf{}%
  \fancyhead[L]{\rmfamily\fontsize{8}{10}\selectfont\bfseries\color{Ink}APCL}%
  \fancyhead[R]{\rmfamily\fontsize{7.5}{9}\selectfont\color{Subtle}Idealized Supersonic RDE}%
  \fancyfoot[R]{\rmfamily\fontsize{8.5}{10}\selectfont\bfseries\color{Cobalt}\thepage}%
}
\fancypagestyle{apclfirst}{%
  \fancyhf{}%
  \fancyfoot[L]{\rmfamily\fontsize{7.6}{9.2}\selectfont\color{Subtle}\textcopyright\ 2026 All rights reserved by the Author and Advanced Propulsion Concepts Lab.}%
  \fancyfoot[R]{\rmfamily\fontsize{8.5}{10}\selectfont\bfseries\color{Cobalt}\thepage}%
}
\newtcolorbox{abstractpanel}{
  enhanced,breakable,boxrule=0pt,colback=Pale,
  borderline west={2.2pt}{0pt}{Cobalt},
  sharp corners,left=13pt,right=12pt,top=11pt,bottom=10pt,
  before skip=0pt,after skip=8pt
}
\newcommand{\minilabel}[1]{{\rmfamily\fontsize{8.5}{10}\selectfont\bfseries\color{Cobalt}\MakeUppercase{#1}}}

\begin{document}
\thispagestyle{apclfirst}

% --- APCL masthead: cropped, original vector wordmark ---
\noindent\begin{minipage}[c]{0.60\linewidth}
  \includegraphics[width=1.69in]{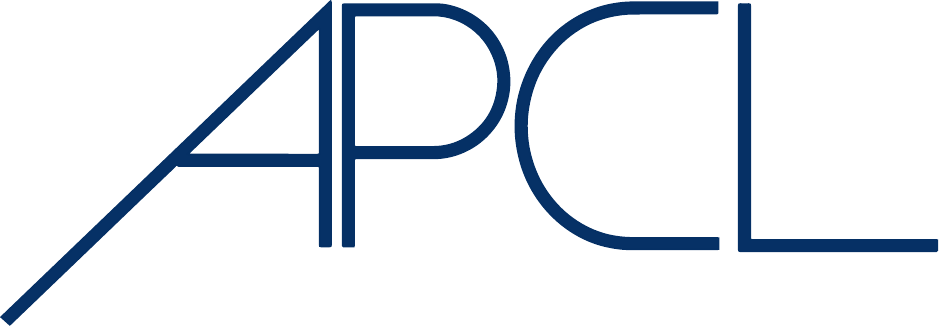}\par\vspace{2pt}
  %{\rmfamily\fontsize{7.5}{9.2}\selectfont\color{Cobalt}\bfseries
   % ADVANCED PROPULSION CONCEPTS LAB\par}
\end{minipage}\hfill\begin{minipage}[c]{0.37\linewidth}
  \raggedleft\rmfamily\fontsize{7.6}{10.2}\selectfont
  \color{Subtle}arXiv\quad$\bullet$\quad \paperdate
\end{minipage}

\vspace{2pt}
\noindent\textcolor{Cobalt}{\rule{1.69in}{1.6pt}}%
\textcolor{SoftRule}{\rule{\dimexpr\linewidth-0.62in\relax}{0.8pt}}
\vspace{10pt}

{\rmfamily\bfseries\fontsize{23}{27}\selectfont\color{Ink}\papertitle\par}
\vspace{5pt}

\vspace{12pt}
{\rmfamily\fontsize{9.9}{12.4}\selectfont\bfseries\color{Ink}\paperauthors\par}
\vspace{4pt}
{\rmfamily\fontsize{8.9}{12.2}\selectfont\color{Subtle}\paperaffiliations\par}
\vspace{12pt}

\begin{abstractpanel}
  \minilabel{Abstract}\par\vspace{3pt}
  The work presented investigates the flow structure and stability of an airbreathing rotating detonation engine (RDE) in an idealized combustor. The presented derivation of a modified pressure cutoff injection scheme allows for supersonic injection in an idealized two-dimensional RDE as a function of a specified design Mach number and the pressure ratio across the injection surface. Numerical simulations are conducted using a compressible reacting flow solver with finite rate chemistry built on an Adaptive Mesh Refinement (AMR) framework. The fuel mixture is set to a stoichiometric hydrogen-air mixture for all simulations. Stable operation occurs for a wide range of inflow conditions with inflow Mach numbers up to 4. The increased total temperature associated with increasing Mach numbers causes burning at the injection surface during the injector recovery process, resulting in unstable operation. Meanwhile, increasing total pressure facilitates sustained supersonic injection. Analysis of the steady flow structure shows a familiar structure with significant changes to the detonation structure when compared with a rocket RDE. With supersonic inflow, an oblique detonation wave (ODW) forms along the injection surface connected to the moving detonation through a slip line. Changes in the oblique detonation wave and normal detonation angle occur with changes in the inflow Mach number, however the overall detonation structure remains similar when supersonic injection occurs.

  \newcommand{\paperkeywords}{Rotating Detonation Engine; Air Breathing; Modeling; Idealized RDE}
  
  % The two metadata lines below belong INSIDE the abstract panel.
  \par\vspace{8pt}
  {\color{SoftRule}\rule{\linewidth}{0.55pt}}\par\vspace{2pt}
  {\rmfamily\fontsize{8.5}{11.5}\selectfont\color{Subtle}%
    \textbf{\textcolor{Ink}{Keywords}}\quad\paperkeywords\par
    \vspace{3pt}
    \textbf{\textcolor{Ink}{Corresponding author}}\quad\papercorrespondence\par
  }
\end{abstractpanel}

\section{Introduction}
In recent years, there has been an increased interest in detonation-based propulsion due to the potential increase in thermodynamic efficiency compared to deflagration-based combustors \cite{Wolanski_Det_Prop}. Detonation-based combustors alleviate the need for traditional compressors, potentially reducing the required number of moving parts by maintaining a geometrically simple design, while achieving high energy density \cite{Braun_LW}. Various engine concepts have been explored to leverage detonation-based propulsion including the pulse detonation engine (PDE) \cite{bussing1994introduction}, oblique detonation engine \cite{abisleiman2025structure}, and rotating detonation engine (RDE) \cite{rankin2017overview}. The rotating detonation engine has been extensively researched as a possible concept for achieving detonation-based propulsion. Studies have focused primarily on the rocket RDE, where fuel and oxidizer are supplied axially through separate plenums, with detailed work being done on full-scale geometries \cite{SuprajFullRDE, Pal_LargeEddy_FullRDE} and simplified 2-D geometries \cite{Ulman2D, Schewer2012_2D}. High-fidelity simulations have provided insight into the role that triple point instabilities play in the propagation of detonation waves and the complex process of stabilizing these waves \cite{moen1986_Cellular, Radulescu_FailureMech}. Studies have also found how plenum conditions affect the injector recovery time and ultimately the number of waves that are formed \cite{RamanNonidealities_FullRDE}. 

With the promising results and successful operation of rocket RDE designs \cite{Sosa_Exp_RDE}, RDEs have been considered for use in an air-breathing engine cycle. Successful operation of ramjet RDEs has been demonstrated \cite{grybko_Ramjet_RDE}, with four distinct combustion modes identified: deflagration, unstarted rotating detonation, started rotating detonation, and pulsed detonation. These operating modes were controlled by varying the back pressure through adjustments to the exit nozzle contraction ratio. Started rotating detonations were achieved at low contraction ratios, whereas pulsed detonations occurred at the highest contraction ratios. Deflagration was observed over moderate-to-high contraction ratios, while the equivalence ratio was found to strongly influence inlet unstart. Further cycle analyses of airbreathing RDEs have indicated that hydrogen–air configurations can achieve specific impulses of up to $3800$ s and support flight at Mach 5 \cite{braun2013airbreathing}. Additionally, investigations into cavity effects in ramjet RDEs have demonstrated enhanced mixing and increased detonation wave speeds with appropriately designed cavities, although inlet unstart remained a persistent challenge \cite{feng2023effects}. Similar to rocket-based RDEs, ramjet RDEs exhibit comparable wave structures because both fuel and air enter the combustor as subsonic streams.

Airbreathing RDEs can also operate in the supersonic regime although this would require modifications to the RDE concept, allowing for high-speed supersonic inflow. Optimal airbreathing RDE performance is predicted to occur while the flow through the isolator remains supersonic through a cycle analysis \cite{braun2013airbreathing}. To realize this design, better understanding of the flow structure observed and the impact on the operational stability is necessary. The feasibility of supersonic-inflow RDE configurations was first explored using a two-dimensional unsteady mathematical model, which demonstrated the formation of transverse detonation waves \cite{zhdan2008mathematical}. Work by Wang et al. showed the feasibility of an RDE with supersonic inflow at Mach 4 by achieving steady operation in an experimental facility \cite{Wang4-0}. Gupta and Schwer presented a numerical simulation of a supersonic RDE (SRDE) with inflow at Mach 2.8 showing a detailed flow structure in the SRDE \cite{Gupta2-8} including the addition of an oblique detonation wave due to the supersonic inflow. Air-breathing RDE operation requires an understanding of the impact have on the inflow properties and feasibility of steady state operation. 

Detailed analysis of a full-scale RDE can become computationally expensive \cite{SuprajFullRDE} and often lacks the resolution required to capture the detonation structure, with few computational cells spanning the induction length, compared to the $40$ cells per induction length that are recommended to resolve the detonation structure \cite{det_conv}. As a result, 2-D simulations have been used to study the stability and performance of RDEs with a pressure cutoff condition \cite{Schewer2012_2D} to model injection across the injection surface, reducing the computational cost. 

The pressure cutoff condition is derived for a rocket RDE where choked injection occurs, $M = 1$, but cannot be extended to supersonic injection. This work will present a modification to the pressure cutoff condition to allow supersonic injection in a 2-D RDE simulation to better understand the stability of high speed air-breathing RDEs. Adaptive mesh refinement and the pressure cutoff condition presented are utilized to analyze the flow structure in an air-breathing RDE and study feasible operating conditions.

\section{Numerical Methods}
The simulations presented are conducted using an in-house compressible reacting flow solver \cite{HMM_Numerics} for the Navier-Stokes equations utilizing adaptive mesh refinement. The reacting compressible Navier-Stokes equations utilized for this study are given by:

\begin{equation}
    \frac{\partial \rho}{\partial t}  + \frac{\partial \rho u_l}{\partial x_l}  = 0,
\end{equation}
\begin{equation}
    \frac{\partial \rho u_j}{\partial t}  + \frac{\partial \rho u_l u_j}{\partial x_l}  = - \frac{\partial p }{\partial x_j}  +  \frac{\partial \tau_{jl} }{\partial x_l},
\end{equation}
\begin{equation}
    \frac{\partial \rho e }{\partial t}  + \frac{\partial u_l \rho h}{\partial x_l} = \frac{\partial}{\partial x_l} \alpha \frac{\partial T}{ \partial x_l} + \frac{\partial \tau_{il} u_i }{\partial x_l},
\end{equation}
\begin{equation}
    \frac{\partial \rho Y_k }{\partial t}  + \frac{\partial u_l \rho Y_k}{\partial x_l} = \frac{\partial}{\partial x_l} \rho D \frac{\partial Y_k}{ \partial x_l} + {\Omega}_k,
\end{equation}
\begin{equation}
    k=1,\ldots,N_S,
\end{equation}

\noindent where $\rho$ is the density of the fluid, $u_j$ is the velocity component in the $j$-th direction, $t$ indicates time and $x_j$ is the spatial location in the $j$-th direction. The viscous stress tensor of the $l$-th component in the $j$-th direction is given by $\tau_{jl}$. \textcolor{black}{The total chemical energy is given by $e$, the total enthalpy is given by $h$, and $\alpha$ is the thermal conductivity of the mixture.}  The temperature and mixture diffusivity are given by $T$, $D$. The chemical source term and mass fraction for the $k$-th species are given by $\Omega_k$ and $Y_k$, respectively. 

Convective fluxes are discretized using the Harten-Lax-van Leer-Contact (HLLC) scheme, while a central differencing scheme is used for the diffusive fluxes. This solver has been used for various detonation and RDE problems in the past \cite{Ulman2D, RamanNonidealities_FullRDE, powers2026impact, abisleiman2025structure, oladipo2025channel, van2024nox}. Finite-rate chemistry is conducted using Cantera \cite{cantera} and a 9-species, 21-reaction chemistry model for hydrogen-air combustion \cite{Hon_Mueller_Mech} to capture varying chemical timescales in the combustor.

The base grid resolution for each simulation is $468$ $\mu$m in the wave-normal direction. Refinement is conducted by tagging the pressure and density gradients to refine the detonation front, oblique shocks, and contact surface. Seven refinement levels are used to tag the detonation front, oblique shock, and contact surface, resulting in a resolution of $3.65$ $\mu$m in the wave normal direction. Previous work by Barway et. al. shows that a minimum of $40$ cells per induction length is required to resolve the detonation structure \cite{det_conv}. The mixture induction length was computed as $183.4$ $\mu m$, using a ZND solution, resulting in $51$ cells per induction length for the presented simulations.

\section{Geometry and Boundary Conditions}

\begin{figure*}[h!]
    \centering
    \includegraphics[width=6.5in]{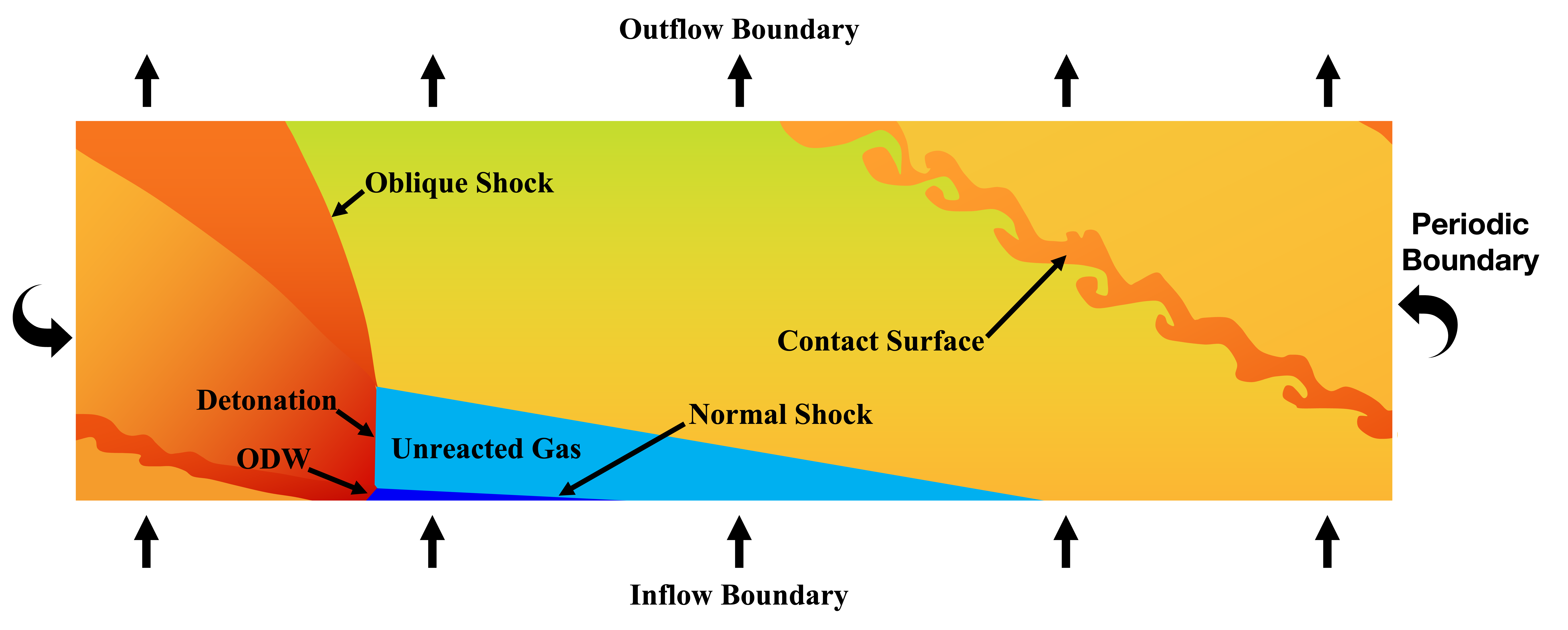}
    \caption{Schematic of geometry and boundary conditions adapted from \cite{Takuma_2021_FullRDE}.
    \label{Geometry}}
\end{figure*}

The geometry utilized in this work, as shown in Fig. \ref{Geometry}, comes from previous work by Ullman, Schewer, and Sato \cite{Ulman2D, Schewer2012_2D, Takuma2018_2DRDE}. The dimensions are $0.4788$ m and $0.14$ m in the x-direction and the y-direction, respectively. This is a linearized RDE where the annulus is unwrapped and periodic boundary conditions are enforced on the left and right boundaries. A supersonic outflow boundary condition is imposed on the top boundary while the bottom boundary utilizes a pressure cutoff condition, derived from a converging nozzle \cite{Schewer2012_2D}, to supply premixed reactants. While capable of modeling injection in a rocket RDE, for an RDE with supersonic inflow, a modified pressure cutoff condition is required. In the following section these equations are derived from nozzle relations across a converging-diverging nozzle. In this formulation, the pressure ratio between the prescribed plenum pressure, $P_{0}$, and the static pressure at the boundary cell, $P_{wall}$, is used to compute the injection properties. The inflow conditions at each boundary cell are determined by the conditions shown below, similar to having a small injector for each cell across the injection surface modeled by a converging-diverging nozzle.

\vspace{-20pt}
\begin{gather*}
    P_{w} > P_0 \rightarrow \text{No injection,} \\
    P_0 > P_{w} > P_{cr1} \rightarrow \text{Subsonic injection,} \\
    P_{cr2} < P_{w} < P_{cr1} \rightarrow \text{Subsonic injection with normal shock,} \\
\end{gather*}
\vspace{-35pt}
\begin{equation}
    P_{w} < P_{cr2} \rightarrow \text{Supersonic injection,}
    \label{BC_Def}
\end{equation}

The two critical pressures, $P_{cr1}$ and $P_{cr2}$, are identified as the nozzle choking pressure and the exit shock pressure (where a normal shock stands precisely at the exit plane), respectively. For each simulation, a design Mach number, $M_D$, and plenum pressure, $P_0$, are prescribed that determine the critical pressures and inflow conditions. The Mach area relation, equation \ref{MachArea}, is utilized to compute $P_{cr1}$ from the design Mach number, $M_D$. The corresponding subsonic solution for the exit Mach number, $M_{cr1}$, is used to compute $P_{cr1}$ with the isentropic equation \ref{PCR1_Def}.

\begin{equation}
    \left(\frac{A}{A^*}\right)^2=\frac{1}{M_D^2}\left[\frac{2}{\gamma+1}\left(1+\frac{\gamma-1}{2} M_D^2\right)\right]^{(\gamma+1) /(\gamma-1)},
    \label{MachArea}
\end{equation}

\begin{equation}
    {{P_{cr1}}}={{P}_0}\left({1}+\frac{\gamma-1}{2} {M_{cr1}}^2\right)^{\frac{-\gamma}{\gamma-1}},
    \label{PCR1_Def}
\end{equation}

To compute $P_{cr2}$, the post normal shock quantities, $M_2$ and $\frac{P_{02}}{P_{01}}$, are found knowing that the pre-shock Mach number, $M_1$, is equal to the design Mach number, $M_D$. Using known pressure ratios, $P_{cr2}$ can be computed as a function of the prescribed plenum pressure and is equal to the exit pressure $P_e$.

\begin{equation}
    {P_{cr2}}={P_0}\frac{P_e}{P_{0_e}} \frac{P_{0_e}}{P_{0_2}} \frac{P_{0_2}}{P_{0_1}} \frac{P_{0_1}}{P_0} = P_e,   \frac{P_{0_e}}{P_{0_2}} = \frac{P_{0_1}}{P_0} = 1,
    \label{PRC2_Def}
\end{equation}

$\frac{P_e}{P_{0_e}}$, defined as the static pressure to total pressure ratio at the exit, is computed using the isentropic equation, equation \ref{ExitPressure}, noting that the flow after the shock is isentropic.

\begin{equation}
    \frac{{P_{e}}}{P_{0_e}}=\left({1}+\frac{\gamma-1}{2} {M_{2}}^2\right)^{\frac{-\gamma}{\gamma-1}},
    \label{ExitPressure}
\end{equation}

With the critical pressures, the three flow regimes are defined, as shown by equation \ref{BC_Def}, and the inflow conditions are calculated as follows.

\subsection{Subsonic Inflow}
Subsonic inflow presents a pressure matched condition, where $P_{inj} = P_{wall}$, and the injection Mach number is found using equation \ref{subsonicMach}.

\begin{equation}
    {M}_{{inj}}=\sqrt{\left[\left(\frac{P_{inj}}{P_0}\right)^{-\frac{\gamma-1}{\gamma}}-1\right] \frac{2}{\gamma-1}},
    \label{subsonicMach}
\end{equation}

\subsection{Subsonic Inflow with Normal Shock}
When a normal shock sits inside the nozzle, the inflow Mach number is found using equation \ref{NormalMach}, as shown by Anderson \cite{Anderson}. Given this is a pressure matched condition, $\frac{P_0 }{P_{inj}}$ is the known pressure ratio and $\frac{A_e^*}{A_e}$ is the nozzle area ratio, also known by the prescribed design Mach number. This analytical solution removed the need to determine the shock location through an iterative solution.

\begin{equation}
    M_{inj}= \sqrt{f(\gamma) \sqrt{\left(\frac{P_0 A_e^*\!}{P_{inj} A_e\!}\right)^2}},
    f(\gamma) = -\frac{1}{\gamma-1}+\sqrt{\frac{1}{(\gamma-1)^2} + \left(\frac{2}{\gamma-1}\right)\left(\frac{2}{\gamma+1}\right)^{\frac{\gamma+1}{\gamma-1}} },
    \label{NormalMach}
\end{equation}
\vspace{-10pt}

\subsection{Supersonic Injection}
At design conditions, the inflow is at the design Mach number, $M_{inj} = M_D$, and the injection pressure can be found using equation \ref{SuperPressure}.

\begin{equation}
    P_{i n j}=P_0\left(1+\frac{\gamma-1}{2} M_D^2\right)^{\frac{-\gamma}{\gamma-1}},
    \label{SuperPressure}
\end{equation}

The effects of over-expanded and under-expanded nozzles are neglected in this boundary condition definition because these effects occur after the injection plane and would not affect the properties at the injection surface. With the injection pressure and Mach number known, the injection temperature and velocity can be found using equations \ref{InjectionTemp} and \ref{InjectionVelocity}, respectively. These values are then used to compute the fluxes across the injection surface.

\begin{equation}
    {T}_{{inj}}={T}_0\left(1+\frac{\gamma-1}{2} M_{{inj}}^2\right)^{-1},
    \label{InjectionTemp}
\end{equation}

\begin{equation}
    U_{inj, y} = \sqrt{\gamma RT_{inj}}*M_D,
    \label{InjectionVelocity}
\end{equation}

\section{Results and Discussions}

\begin{figure*}[h!]
    \centering
    \includegraphics[width=6.5in]{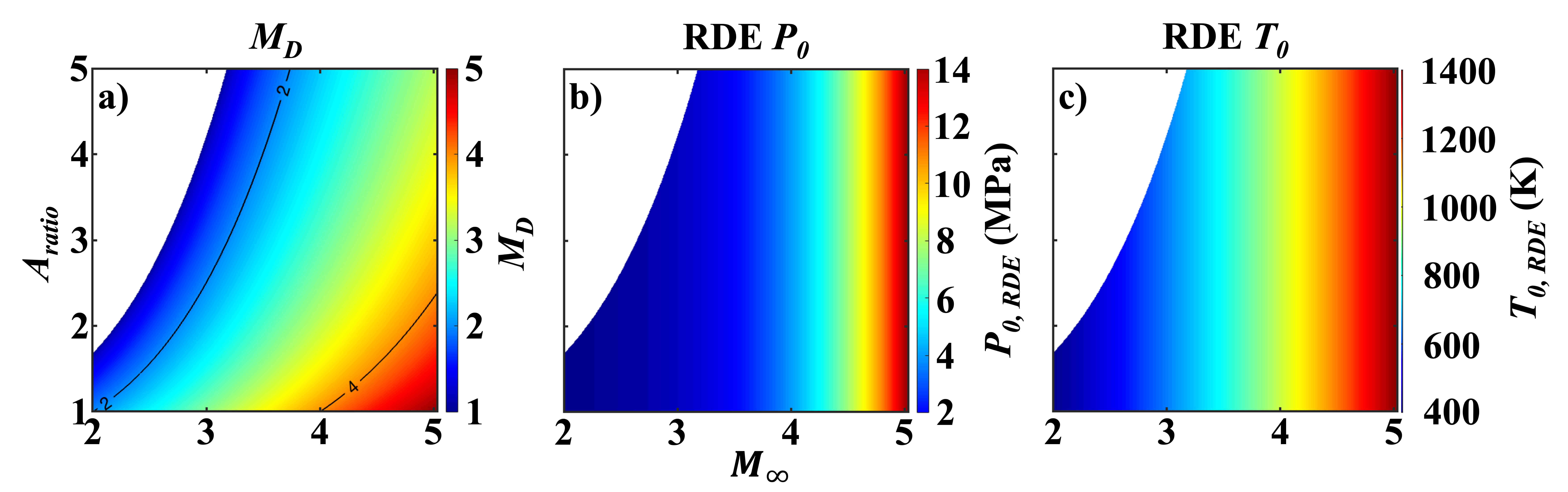}
    \caption{RDE Conditions as a function of freestream Mach number and inlet area ratio:
    a) Design Mach number for RDE, b) RDE plenum pressure, c) RDE plenum temperature.
    \label{RDEConditionPlot}}
\end{figure*}

To determine appropriate plenum conditions and design Mach number for an air-breathing RDE, an isentropic inlet takes freestream conditions to the design Mach number. Figure \ref{RDEConditionPlot} shows the RDE design conditions as a function of the freestream Mach number and the intake area ratio. Plenum conditions are only a function of the freestream Mach number and result in the plenum properties increasing with the freestream Mach number.

\begin{figure}[h!]
    \centering
    \includegraphics[width = 4.00in, trim=0 170 0 0, clip]{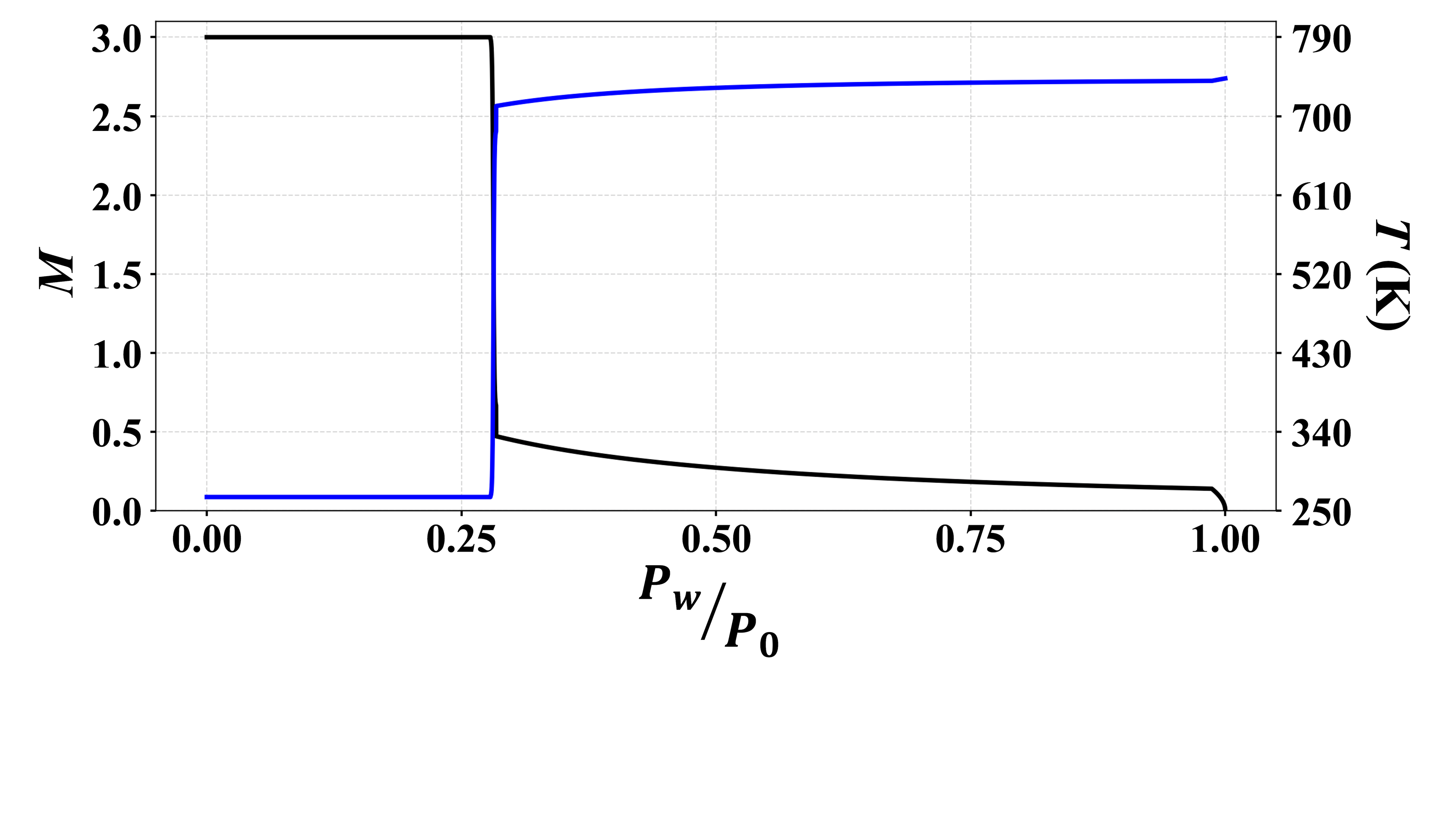}
    \caption{Mach number and Temperature profile as a function of pressure ratio imposed at the injection surface for Case 3: \protect\brightbluecircle~$\rightarrow $ Temperature, \protect\blackcircle~$\rightarrow $ Mach number.}
    \label{TempProfile}
\end{figure}

This results in high plenum pressures and temperatures as the flight speed increases, which presents both benefits and challenges. As the detonation wave passes over an injector, the pressure increases at the injection surface and causes the injector to be momentarily blocked until the pressure is relieved. Figure \ref{TempProfile} shows the injection temperature profile for a Mach = $3$ inflow condition. The discontinuity at $\frac{P_w}{P_0} \approx 0.29$ corresponds with $P_{cr2}$. High pressure ratios above $P_{cr2}$ occur during the recovery process and results in injection temperatures within $95.7 \%$ of the freestream total temperature. 

Normal shocks in the nozzle enforce subsonic injection during the recovery process when pressure ratios are above $P_{cr2}$ as seen in fig. \ref{TempProfile}. This results in sustained injection at high temperature that approaches the freestream total temperature, significantly above the designed injection conditions. For high freestream Mach numbers, the injection temperature during the recovery process can exceed the auto-ignition temperature of the reactants. This results in burning at the injection site and could act as a spark in the domain leading to the formation of additional counter-rotating waves. As a result, there is a limitation to the allowable freestream Mach number as the plenum conditions should remain below the auto-ignition temperature of the fuel. For a stoichiometric hydrogen-air mixture, this limits the freestream Mach number to $4.03$. Lower area ratio intakes could be beneficial at higher flight velocities where the RDE design Mach number would be allowed to approach the freestream Mach number. From these plots, various test cases were identified, highlighted in Table \ref{TestCases}.

\begin{table}[htbp]
    \centering    \renewcommand{\arraystretch}{1.2}
    \setlength{\tabcolsep}{8pt}

    \begin{tabular}{cccccc}
        \toprule
        \textbf{ } &
        \textbf{$M_{D}$} & \textbf{$M_\infty$} & \textbf{$P_0$} (kPa)&
        \textbf{$T_0$} (K)& \textbf{$A_{r}$} \\
        \midrule
        Case 1 & 2.0 & 2.760 & 646 & 558 & 2.00 \\
        Case 2 & 2.5 & 3.095 & 1122 & 651 & 1.75 \\
        Case 3 & 3.0 & 3.421 & 1789 & 743 & 1.50 \\
        Case 4 & 3.5 & 3.749 & 2864 & 850 & 1.25 \\
        Case 5 & 4.0 & 4.000 & 3984 & 935 & 1.00 \\
        \bottomrule
    \end{tabular}
    \caption{Test cases identified assuming isentropic compression.}
    \label{TestCases}
    \vspace{-10pt}
\end{table}

%2.0  & 2.7604 & 646      & 558    & 2.00   \\
%2.5  & 3.0950 & 1122     & 651    & 1.75   \\
%3.5  & 3.7489 & 2864     & 850    & 1.25   \\
%4.0  & 4.0000 & 3984     & 935    & 1.00  

Figure \ref{TempCFD} below shows the steady state operation of a linearized RDE with $M_D = 3$. Supersonic flow in an RDE combustor presents a similar flow structure to the rocket RDE \cite{Ulman2D, Schewer2012_2D} with the addition of an oblique detonation wave (ODW) that sits along the injection surface. This ODW is connected to a normal detonation wave via a slip line shown on the right contour in fig. \ref{TempCFD}. The unreacted gas section is also divided by a normal shock that contains supersonic unreacted gas and subsonic unreacted gas. Lastly, coming from the ODW, there is an additional contact surface that expands downstream towards the outlet, eventually interacting with the oblique shock originating from the detonation wave. These results are in agreement with those found by Schwer and Gupta \cite{Gupta2-8}. 

\begin{figure*}[h!]
    \centering
    \includegraphics[width = 6.5in]{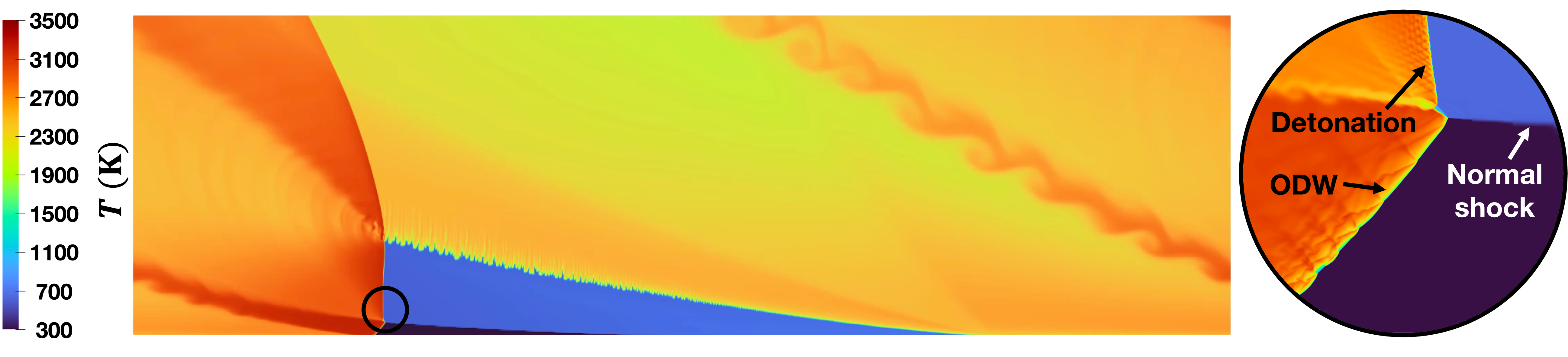}
    \caption{Steady state temperature contours with injection at Mach $= 3$ of premixed, stoichiometric hydrogen-air including zoomed in view of detonation structure (right).}
    \label{TempCFD}
\end{figure*}

\begin{figure}[h!]
    \centering
    \includegraphics[width = 4.00in]{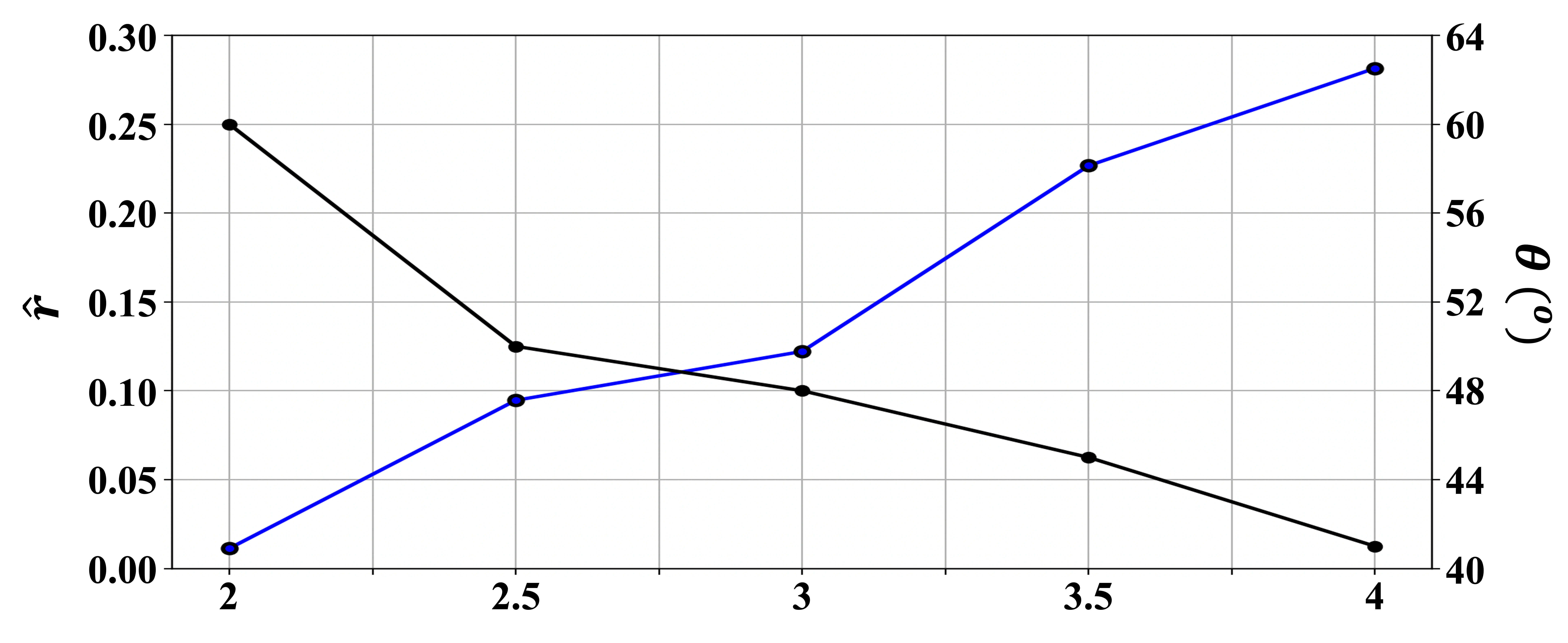}
    \caption{Unburned reactant infill ratio and ODW inclination angle, measured with respect to the injection boundary, as a function of Mach number: \protect\brightbluecircle~$\rightarrow \hat{r}$, \protect\blackcircle~$\rightarrow \theta$.}
    \label{Inclination_Infill}
\end{figure}

The formation of the oblique detonation wave is a result of the increased axial momentum of the inflow with supersonic injection. Analyzing the detonation from the wave frame, there is an axial momentum component from the injectors and an azimuthal component due to the wave propagation. Behind the wave the flow must be parallel to the injection surface resulting in the inclination of the wave towards the injectors to acount for the increase axial momentum. In a rocket RDE, the axial momentum from injection is small with respect to the azimuthal component, resulting in a small inclination of the detonation towards the injectors as observed in numerous studies \cite{Meadows_2DRDE, nakagami2017experimental, powers2026impact}. Supersonic inflow increases the axial momentum and results in the detonation wave tilting further towards the injectors, forming an ODW. Figure \ref{Inclination_Infill} shows that at higher design Mach numbers the ODW tilts further into the injectors due to the increased axial momentum. At Mach $2$ the ODW angle is $60^o$ while reaching $41^o$ at a inflow Mach number equal to $4$. 

All test cases, presented in table \ref{TestCases}, resulted in a similar flow structure as seen in Fig. \ref{TempCFD}, with the formation of an ODW and a normal shock separating the non-reacted flow with supersonic and subsonic reactants. As the freestream Mach number increases, the oblique detonation wave grows and occupies a larger fraction of the infill height. The infill ratio, $\hat r$, is defined as the ratio between the ODW height and the infill height, $\frac{H_{ODW}}{H_{Infill}}$. Figure \ref{Inclination_Infill} shows the infill ratio as a function of the design Mach number where at $M=2$ a small fraction, about $1 \%$, of the infill height is occupied by the ODW. At a design Mach number of $4$, the ODW occupied $\approx 28 \%$ of the infill height. 

Raising the inflow Mach number leads to a rise in total pressure, since it scales proportionally with $M_d^{2 \frac{\gamma}{\gamma-1}}$. The resulting higher plenum pressure enables the injection to recover towards design conditions more rapidly, which helps maintain supersonic injection. Although the post-detonation pressure is similar across cases ($\approx$ $15$ atm), the injection recovery length scales differ substantially: the Mach $2$ case only returns to design conditions about $0.05$m upstream of the detonation, whereas the Mach $4$ case reaches design conditions roughly $0.18$m upstream. Cases $2$ and $3$ attain the target Mach number at about $0.14$ and $0.2$m upstream of the detonation, respectively. Even though the recovery time keeps decreasing with Mach number, the incremental improvement becomes less significant. This behavior is attributed to the reduction of $\frac{P_{cr2}}{P_0}$ as Mach number increases. Nevertheless, up to Mach $4$ the increase in plenum pressure remains large enough to offset this effect.

\begin{figure}[h!] 
    \centering \includegraphics[width = 4.00in]{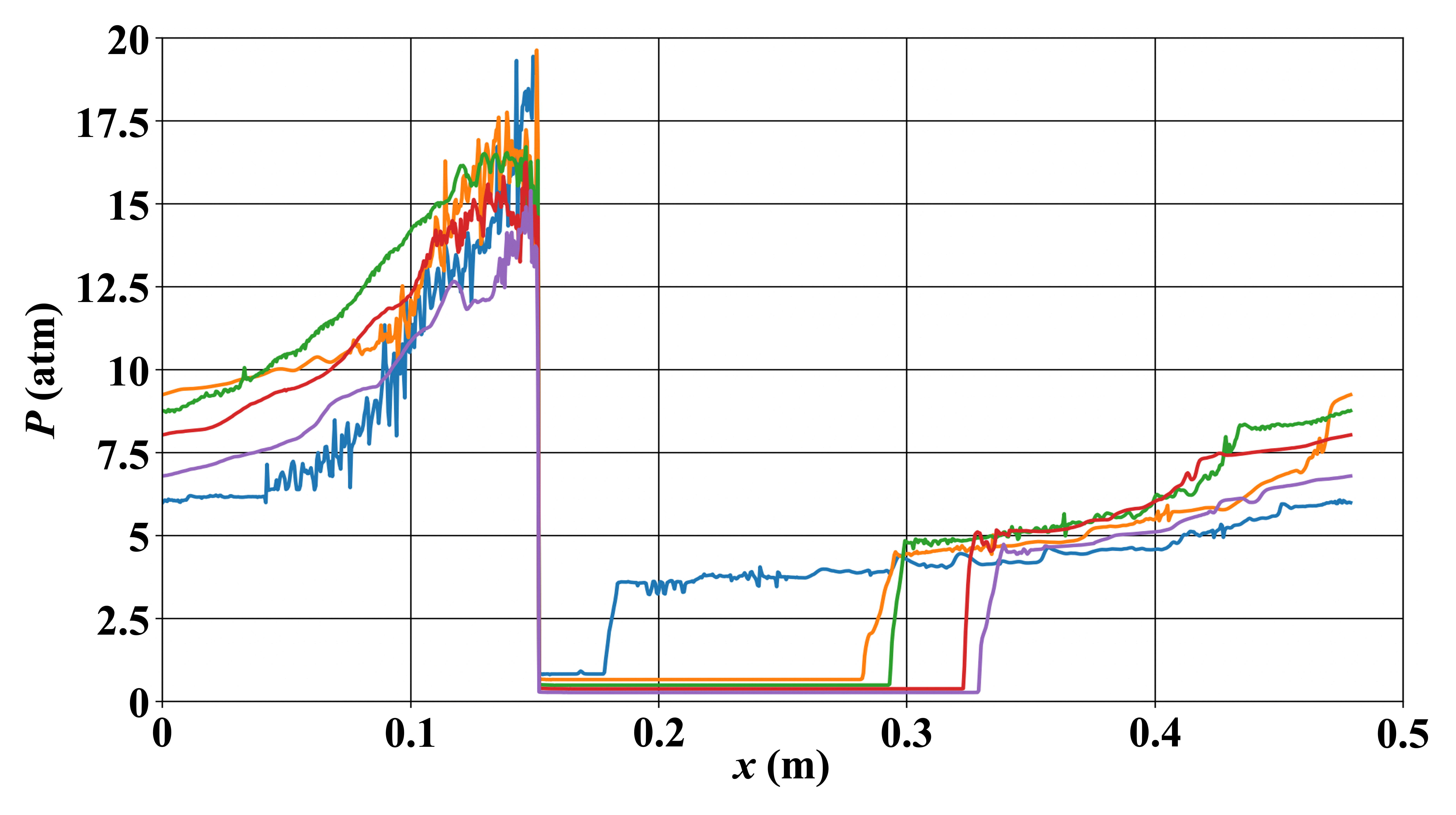} 
    \caption{Pressure profiles near injection surface, $Y = 3.25 \times 10^{-4}$ m, at varying $M_d$
    (\protect\bluecircle~$\rightarrow M=2$, \protect\orangecircle~$\rightarrow M=2.5$,
    \protect\greencircle~$\rightarrow M=3$, \protect\redcircle~$\rightarrow M=3.5$,
    \protect\purplecircle~$\rightarrow M=4$).}
    \label{InjectorRecovery} 
\end{figure}

\begin{figure}[h!]
    \centering
    \includegraphics[width=4.00in]{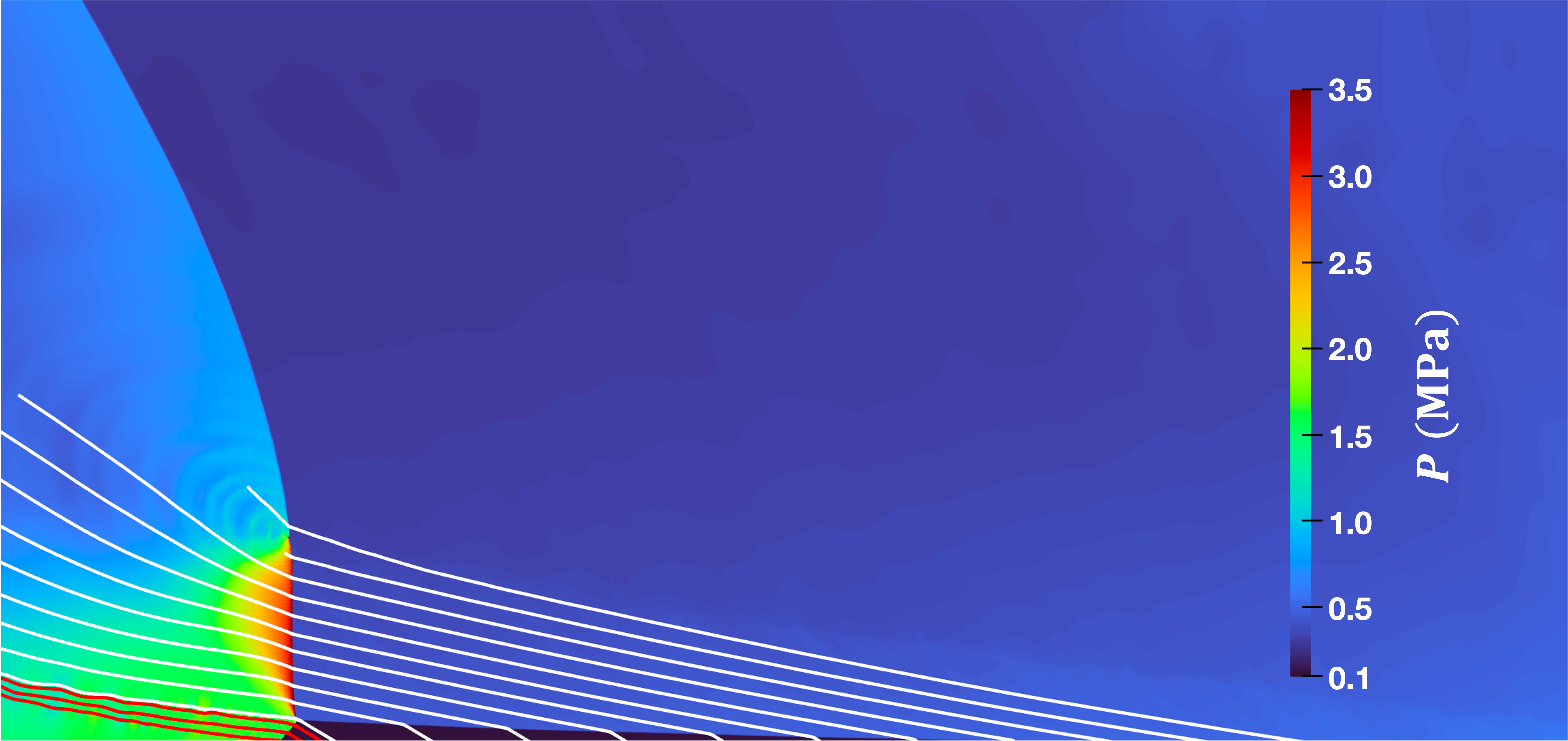}
    \caption{Collection of streamlines originating along the injection surface: \protect\redcircle $\rightarrow$ processed by oblique detonation, \protect\whitecircle $\rightarrow$ processed by normal detonation}
    \label{Streamlines}
\end{figure}

The existence of a normal shock in the unburnt reactants introduces an alternative thermodynamic trajectory for the reactants. Streamlines from the inlet, shown in fig. \ref{Streamlines}, indicate that reactants injected close to the detonations are primarily processed by the oblique detonation wave. In contrast, reactants introduced farther upstream encounter the normal shock before subsequently passing through the detonation. The resulting rise in thermodynamic state can trigger reactions upstream of the detonations, commonly referred to as parasitic deflagration. This may lower combustor efficiency while also encouraging unstable wave propagation. At inflow Mach numbers above $4$, the normal-shock-induced increase in the upstream thermodynamic state is what leads to unstable operation.

\section{Conclusion}
In this study, the pressure cutoff boundary conditions typically applied in 2-D RDE simulations are modified to enable supersonic injection, and then employed to assess the viability of an air-breathing RDE, along with the examination of the combustor flowfield. The findings indicate that sustained, steady air-breathing RDE operation is achievable, with an overall structure comparable to that of a rocket-based RDE. Key distinctions are the emergence of an oblique detonation wave and a normal shock that divides the infill region into supersonic and subsonic portions. Stable operation was demonstrated for injection Mach numbers up to 4. An evaluation of plenum conditions as a function of freestream flight velocity shows that, at high flight speed, recovery raises the injection temperature, leading to combustion at the injectors and the formation of additional counter-rotating waves. Increasing the inflow total pressure, raises the fraction of fuel processed by the oblique detonation up to $28 \%$, whereas increasing the inflow Mach number increases the oblique detonation angle to offset the higher axial momentum, with wave angles of $41 \%$ reported at Mach $4$.

\section*{CrediT authorship contribution statement}

{\bf Ruben A. Fernandez}: Writing - Original draft, visualization, investigation, data analysis, methodology, formal analysis, data collection, conceptualization, software. 

{\bf Venkat Raman}: Writing - Review and editing, supervision, software, resources, project administration, funding acquisition, formal analysis, conceptualization.

\section*{Declaration of competing interest}

The authors declare that they have no known competing financial interests or personal relationships that could have appeared to influence the work reported in this paper.

\section*{Acknowledgements} The research is supported by the Air Force Office of Scientific Research (AFOSR) under Grant No. FA9550-24-1-0017, with Dr. Chiping Li as Program Officer. The computational resources were provided by the U.S. Department of Defense High-Performance Computing Modernization Program (DoD HPCMP).

% Numerical citations [1], [2] and reference list ordered by first citation.
\bibliographystyle{unsrtnat}
\bibliography{references}
\end{document}